\documentclass[preprint2]{aastex6}

\usepackage{graphicx}
\usepackage{amsmath, amssymb}
\usepackage{multirow}
\usepackage[version=3]{mhchem}

\usepackage{natbib}
\bibpunct{(}{)}{;}{a}{}{,}

\graphicspath{{./}}

\newcommand{\reffig}[1]{Fig.~\ref{#1}}
\newcommand{\reftab}[1]{Table~\ref{#1}}
\newcommand{\refsec}[1]{Section~\ref{#1}}

\newcommand{\Ti}{$1_{10}-1_{01}$}
\newcommand{\Tii}{$1_{11}-0_{00}$}
\newcommand{\Tiii}{$3_{12}-2_{21}$}
\newcommand{\Tiv}{$3_{12}-3_{03}$}

\begin{document}

\title{Survey of cold water lines in protoplanetary disks: indications of systematic volatile depletion}

\author{
Fujun~Du\altaffilmark{1},
Edwin~Anthony~Bergin\altaffilmark{1},
Michiel~Hogerheijde\altaffilmark{2},
Ewine~F.~van~Dishoeck\altaffilmark{2,3},
Geoff~Blake\altaffilmark{4},
Simon~Bruderer\altaffilmark{3},
Ilse~Cleeves\altaffilmark{5},
Carsten~Dominik\altaffilmark{6},
Davide~Fedele\altaffilmark{7},
Dariusz~C.~Lis\altaffilmark{8,9},
Gary~Melnick\altaffilmark{5},
David~Neufeld\altaffilmark{10},
John~Pearson\altaffilmark{11},
Umut~Y{\i}ld{\i}z\altaffilmark{11}
}
\altaffiltext{1}{Department of Astronomy, University of Michigan, 311 West Hall, 1085 S. University Ave, Ann Arbor, MI 48109, USA}
\altaffiltext{2}{Leiden Observatory, Leiden University, P.O. Box 9513, 2300 RA, Leiden, The Netherlands}
\altaffiltext{3}{Max Planck Institut f\"ur Extraterrestrische Physik, Giessenbachstrasse 1, 85748, Garching, Germany}
\altaffiltext{4}{Division of Geological \& Planetary Sciences, MC 150-21, California Institute of Technology, 1200 E California Blvd, Pasadena, CA 91125}
\altaffiltext{5}{Harvard-Smithsonian Center for Astrophysics, 60 Garden Street, Cambridge, MA 02138, SA}
\altaffiltext{6}{Astronomical institute Anton Pannekoek, University of Amsterdam, Science Park 904, 1098 XH, Amsterdam, The Netherlands}
\altaffiltext{7}{INAF/Osservatorio Astrofisico di Arcetri, Largo E. Fermi 5, 50125, Firenze, Italy}
\altaffiltext{8}{LERMA, Observatoire de Paris, PSL Research University, CNRS, Sorbonne Universites, UPMC Univ. Paris 06, F-75014, Paris, France}
\altaffiltext{9}{Cahill Center for Astronomy and Astrophysics 301-17, California Institute of Technology, Pasadena, CA 91125, USA}
\altaffiltext{10}{Department of Physics and Astronomy, Johns Hopkins University, 3400 North Charles Street, Baltimore, MD 21218, USA}
\altaffiltext{11}{Jet Propulsion Laboratory, California Institute of Technology, Pasadena, CA 91109, USA}


\begin{abstract}
We performed very deep searches for 2 ground-state water transitions in 13 protoplanetary disks with the \textsl{HIFI} instrument on-board the \textsl{Herschel Space Observatory}, with integration times up to 12 hours per line.  Two other water transitions that sample warmer gas were also searched for with shallower integrations.  The detection rate is low, and the upper limits provided by the observations are generally much lower than predictions of thermo-chemical models with canonical inputs.  One ground-state transition is newly detected in the stacked spectrum of AA~Tau, DM~Tau, LkCa~15, and MWC~480.  We run a grid of models to show that the abundance of gas-phase oxygen needs to be reduced by a factor of at least ${\sim}100$ to be consistent with the observational upper limits (and positive detections) if a dust-to-gas mass ratio of 0.01 were to be assumed.  As a continuation of previous ideas, we propose that the underlying reason for the depletion of oxygen (hence the low detection rate) is the freeze-out of volatiles such as water and CO onto dust grains followed by grain growth and settling/migration, which permanently removes these gas-phase molecules from the emissive upper layers of the outer disk.  Such depletion of volatiles is likely ubiquitous among different disks, though not necessarily to the same degree.  The volatiles might be returned back to the gas phase in the inner disk (${\lesssim}15$~AU), which is consistent with current constraints.  Comparison with studies on disk dispersal due to photoevaporation indicates that the timescale for volatile depletion is shorter than that of photoevaporation.
\end{abstract}

\keywords{astrochemistry --- circumstellar matter --- molecular processes ---
planetary systems --- planet-disk interactions --- planets and satellites:
atmospheres}

\section{Introduction}
\label{secIntroduction}

Dust evolution in protoplanetary disks is inevitably coupled with gas evolution and gas-phase chemistry.  The dust distribution determines the UV field of the disk, which strongly affects the gas temperature \citep{Glassgold2004,Woitke2009} and photo-reaction (photodissociation and photodesorption) rates \citep{Bergin2003,vanDishoeck2006,Bruderer2009a,Oberg2009a}.  Dust grains also act as a sink of gas-phase molecules, because molecules can freeze out onto the dust grains.  The freeze-out process changes the composition of dust particles, and can potentially alter the sticking coefficient of dust coagulation \citep{Machida2010,Sato2016}.  Among all the gas-phase species, water is of great interest, not just because of its importance for the origin of water on planets and the possibility for life elsewhere in the universe  (see review by \citealt{vanDishoeck2014}), but also because (1) it is the major carrier of the third most abundant element, oxygen, (2) it is an important coolant \citep{Karsai2013a,Nisini2010}, and (3) it may shield other molecules from Lyman~\(\alpha\) photons in some parts of the disk \citep{Bethell2011a,Du2014,Adamkovics2014}.

We have surveyed 4 water lines in 13 protoplanetary disks with the \textsl{HIFI} instrument of the \textsl{Herschel Space Telescope}\footnote{Herschel is an ESA space observatory with science instruments provided by European-led Principal Investigator consortia and with important participation from NASA.} (\reftab{tabObsInfo}).  Two of these lines (the ground state lines) have \emph{very deep} integrations, and two higher-lying lines have moderately deep integration.  All the spectra are resolved in velocity.  The disks are heterogeneous in nature, with a range of disk masses (in dust) and stellar types.  This survey is mostly sensitive to cold ($T\lesssim20$~K) water vapor from the outer part of the disk (for a quantitative view, see \refsec{subsecDiskSize}, where we show that the inner few astronomical units contribute less than 10\% to the transitions studied in this survey).  In contrast, starting with Carr et al. (2004), many papers have probed the warm--hot water reservoir in the inner few AU of disks via infrared lines (e.g., \citealt{Carr2008,Carr2011}; \citealt{Pontoppidan2010a,Pontoppidan2010b}; \citealt{Salyk2008,Salyk2011,Salyk2015}; \citealt{Banzatti2012,Banzatti2015}; and \citealt{Blevins2016}).

The major finding is that while a few sources show positive detections, most of them have no signal at the current noise level, even with very deep integrations, consistent with our earlier Herschel-HIFI results \citep{Bergin2010, Hogerheijde2011}.  We model these water lines with a thermo-chemical code described below, and find that the model predictions tend to be much higher than the observed upper limits, with reasonable and canonical parameters for the sources.  This phenomenologically resembles other studies on carbon-bearing species including CO and atomic carbon in HD~100546, TW~Hya, and DM~Tau \citep{Bruderer2012,Favre2013,Cleeves2015,Kama2016,Schwarz16,Bergin2016}.  Similar to previous conclusions \citep{Du2015,Bergin2016,Kama2016a}, a natural explanation for the disparity between the models and the observations is that volatile elements such as oxygen freeze out onto dust grains and never return back to the gas phase due to dust growth (possibly into multi-kilometer-sized bodies) and settling/migration.

This paper is organized as follows.  \refsec{secSurvey} describes details of the water line survey, \refsec{secModel} presents a grid of models proposed as an effort to match the observations, and in \refsec{secResults} we compare the model results with observed upper limits and detections.  In \refsec{secDiscussion} we discuss the implications of the comparison between the models and observations, and finally in \refsec{secSummary} we summarize the main findings.

\section{The water line survey}
\label{secSurvey}

The observations presented in this paper were obtained with the Heterodyne
Instrument for the Far-Infrared \citep[HIFI;][]{degraauw2010, roelfsema2012}
on-board the \textsl{Herschel Space Observatory} \citep{pilbratt2010} as part of the
Guaranteed Time Key Project ``Water in Star-forming Regions with Herschel''
\citep[WISH;][]{vandishoeck2011} and two open-time programs (PI: Hogerheijde).
With its FWHM (Full Width at Half Maximum) beam of $18''$--$38''$ at the observing frequencies, \textsl{Herschel} covers
the entire disk of all sources.  \reftab{tabObsInfo} gives an overview of the
observed sources, lines, dates, OBS~ID identifiers and observing times $t_{\rm
obs}$.  On-source integration times range between 35\% and 45\% of $t_{\rm
obs}$, with a larger fraction spent on-source for larger $t_{\rm obs}$; the
remainder of $t_{\rm obs}$ was spent on the off positions and telescope
overheads.  Three frequency settings cover more than one line: the setting at
556.9 GHz contains the NH$_3$ $1_0$--$0_0$ line at 572.49817 GHz, the setting
at 1113.3 GHz contains the $^{13}$CO 10--9 line at 1101.3496594 GHz, and the
1153.1 GHz setting covers the $^{12}$CO 10--9 line at 1151.985452 GHz.
Detections of H$_2$O $1_{10}$--$1_{01}$ and $1_{11}$--$0_{00}$ and
$^{12}$CO 10--9 toward TW Hya and HD~100546, and NH$_3$ $1_0$--$0_0$ toward TW Hya
only, are presented elsewhere \citep[][Hogerheijde et al. in
prep.]{Hogerheijde2011, fedele2013, Salinas2016, Fedele2016} (the \Ti{} line of HD~100546 can be seen in Fig.~3 of \citealt{vanDishoeck2014}).  An earlier
report on DM~Tau containing only a fraction of the data presented here can be
found in \citet{Bergin2010}.
In appendix \ref{appendixObsFigs} we show the observed spectra of all the sources and lines that have been surveyed (the CO data for HD~100546 have been published before in \citealt{Fedele2013a}).
Here we focus on the non-detections (upper
limits) of the full data set, with the exception that we also report a new
detection of the \Ti{} line in the stacked spectrum of AA~Tau, DM~Tau, LkCa~15,
and MWC~480 (\refsec{secStackDetec}).
 
We performed very deep observations for the \Ti{} and \Tii{} transitions in four sources (AA~Tau, DM~Tau, LkCa~15, MWC~480).  The integration times were motivated by the TW~Hya detection.  Thus, scaled for distance, if other sources were as strong as TW~Hya, we should be able to detect the two lines.  To account for the possibility that the ground state lines might not be well matched to the excitation state of the disk gas, we also observed higher transitions (with less integration times).

For all observations we used dual beam switch mode with a $3'$ throw.  The spectra were recorded with the Wide-Band Spectrometer (WBS) and High Resolution Spectrometer (HRS) with respective velocity resolutions of 0.59 and 0.13 km~s$^{-1}$ around 550 GHz, and 0.3 and 0.067 km~s$^{-1}$ around 1100 GHz.  The data are processed with various HIPE (Herschel Interactive Processing Environment) versions (4.0--12.1.0), which all give consistent results, and are further analyzed with the CLASS software package\footnote{See \url{http://www.iram.fr/IRAMFR/GILDAS}}.  HIFI measures the vertical and horizontal polarizations separately.  We averaged the two polarization signals together, after verifying their consistency, and only report the upper limits based on the WBS data, because of their lower noise.  Intensities on the main-beam antenna temperature scale follow from the in-orbit calibrated $T_A^*$ antenna-temperature scale using main beam efficiencies $\eta_{\rm mb}$=0.64 around 550 GHz and 0.63 around 1100 GHz\footnote{HIFI-ICC-RP-2014-001 at \url{http://herschel.esac.esa.int/twiki/bin/view/Public/HifiCalibrationWeb}}.  Finally, linear spectral baselines are subtracted in a ${\sim}40$ km~s$^{-1}$ range around the line frequencies, and rms noise levels are extracted (\reftab{tabObsData}).  The rms of the integrated intensities (when not detected) are calculated from the FWHM of the low-$J$ CO lines and the per-channel noise.  Reported upper limits are 3$\sigma$ where $\sigma= 1.2 \times \text{rms}$, with the extra factor 1.2 taking into account an estimated 20\% flux calibration uncertainty.

\begin{deluxetable*}{rrrrrr}
\tablecolumns{6}
\tablewidth{0pc}
\tablecaption{Overview of the water observations
\label{tabObsInfo}}
\tablehead{
 & \colhead{Frequency} & & & & \colhead{$t_{\rm obs}$}\\
\colhead{Line} & \colhead{(GHz)} & \colhead{Source} & \colhead{Date} & \colhead{OBS~ID} & \colhead{(s)}
}
\startdata
$1_{11}$--$0_{00}$ & 1113.342964 & AA Tau   & 2012-09-04 & 1342250602 & 40542.4\\
                  &             &           & 2013-02-26 & 1342266479 & 20196.4\\
                  &             & DM Tau    & 2010-03-04 & 1342191652 & 43235.3\\
                  &             &           & 2012-08-30 & 1342250453 & 36576.4\\
                  &             &           & 2012-08-31 & 1342250454 & 36340.4\\
                  &             & HD 163296 & 2012-10-17 & 1342253594 & 19031.4\\
                  &             &           & 2012-10-18 & 1342253595 & 40285.4\\
$1_{10}$--$1_{01}$ &  556.936002 & AA Tau   & 2012-03-27 & 1342242495 & 30270.0\\
                  &             &           & 2012-03-27 & 1342242496 & 29286.5\\
                  &             & DM Tau    & 2010-03-20 & 1342192365 & 23903.9\\
                  &             &           & 2012-08-29 & 1342250427 & 17863.9\\
                  &             &           & 2012-09-06 & 1342250687 & 19244.9\\
                  &             & HD 163296 & 2010-03-21 & 1342192516 &  1914.4\\
                  &             & LkCa 15   & 2010-08-31 & 1342204003 &  24044.9\\
                  &             & MWC 480   & 2010-09-01 & 1342204004 &  23888.9\\
                  &             & AS 209    & 2010-03-21 & 1342192518 &   2087.4\\
                  &             & BP Tau    & 2010-03-21 & 1342192523 &   2421.4\\
                  &             & GG Tau    & 2010-08-19 & 1342203193 &   1914.4\\
                  &             & GM Aur    & 2010-08-19 & 1342203209 &   1981.4\\
                  &             & IM Lup    & 2011-02-15 & 1342214336 &   2510.4\\
                  &             & MWC 758   & 2010-04-11 & 1342194502 &   2119.4\\
                  &             & T Cha     & 2010-04-12 & 1342194535 &   2849.4\\
$3_{12}$--$2_{21}$ & 1153.126822 & AA Tau   & 2012-08-15 & 1342249596 & 5296.5\\
                  &             & DM Tau    & 2010-08-20 & 1342203259 &  2263.3\\
                  &             &           & 2012-08-15 & 1342249597 &  2923.7\\
                  &             & LkCa 15   & 2010-08-20 & 1342203257 &   2263.3\\
                  &             & MWC 480   & 2011-04-01 & 1342217734 &   2452.4\\
                  &             & TW Hya    & 2010-12-02 & 1342210733 &   3179.4\\
                  &             & HD 100546 & 2011-12-30 & 1342235779 &   5358.5\\
$3_{12}$--$3_{03}$ & 1097.364791 & DM Tau   & 2010-09-02 & 1342203941 & 14911.7\\
                  &             & LkCa 15   & 2010-09-02 & 1342203939 &  14968.7\\
                  &             & MWC 480   & 2010-09-02 & 1342203936 &  15104.7\\
                  &             & TW Hya    & 2010-06-09 & 1342197986 &  15121.7\\
\enddata
\end{deluxetable*}

\begin{table*}[htbp]
\centering
\caption{Upper limits of the water $1_{10}-1_{01}$, $1_{11}-0_{00}$,
$3_{12}-2_{21}$, and $3_{12}-3_{03}$ lines, together with other basic
parameters of each source.  The quoted upper limits are based on the WBS
measurements.  The values in the upper limit column with an error bar (in boldface) are positive detections.}
\label{tabObsData}
\begin{tabular}{l c r c c c c c c c c c}
\hline\hline
Source      &  Line            &  Upper Limit           &  FWHM          & $M_\text{star}$ & $T_\text{eff}$  & $L_\star$   & SpT     & $M_\text{disk}$\tablenotemark{a} & $d$  & $r_\text{out}$  & Ref             \\  
            &                  &  (mK km s$^{-1}$)      &  (km s$^{-1}$) & ($M_\odot$)     & (K)             & ($L\odot$)  &         & $M_\odot$       & (pc) & (AU)            &                 \\  
\hline                                                                                                                                                                                                %
TW~Hya      & $1_{11}-0_{00}$  &  $\mathbf{54.5\pm3.7}$\tablenotemark{b}   & 1.3            & 0.8             & 4100            & 0.28        & K7      & 0.05            & 51   & 215             & (4,5)           \\
            & $1_{10}-1_{01}$  &  $\mathbf{32.5\pm1.4}$\tablenotemark{b}   & 1.3            &                 &                 &             &         &                 &      &                 &                 \\
            & $3_{12}-2_{21}$  &  116.0                 & 1.0            &                 &                 &             &         &                 &      &                 &                 \\
            & $3_{12}-3_{03}$  &  13.4                  & 1.0            &                 &                 &             &         &                 &      &                 &                 \\  
\hline                                                                                                                                                                                                %
HD~100546   & $1_{11}-0_{00}$  &  $\mathbf{247.0\pm4.1}$\tablenotemark{b}  & 7.3            & 2.4             & 10471           & 32.4        & B9Vne   & 0.005           & 103  & 500             & (6,17,28)       \\
            & $1_{10}-1_{01}$  &  $\mathbf{163.0\pm3.0}$\tablenotemark{b}  & 6.6            &                 &                 &             &         &                 &      &                 &       \\
            & $3_{12}-2_{21}$  &  895.0                 & 6.5            &                 &                 &             &         &                 &      &                 &                 \\
\hline                                                                                                                                                                                                %
AA~Tau      & $1_{11}-0_{00}$  &  70.5                  & 4.0            & 0.76            & 4060            & 0.8         & K7      & 0.02            & 140  & 160             & (12,27)         \\  
            & $1_{10}-1_{01}$  &  19.5                  & 4.0            &                 &                 &             &         &                 &      &                 &                 \\  
            & $3_{12}-2_{21}$  & 583.0                  & 4.0            &                 &                 &             &         &                 &      &                 &                 \\  
\hline                                                                                                                                                                                             %
DM~Tau      & $1_{11}-0_{00}$  &  20.5                  & 1.5            & 0.65            & 3705            & 0.25        & M1      & 0.025           & 140  & 750             & (3,9,13,23,29)  \\  
            & $1_{10}-1_{01}$  &   7.2                  & 1.5            &                 &                 &             &         &                 &      &                 &                 \\  
            & $3_{12}-2_{21}$  & 336.0                  & 1.5            &                 &                 &             &         &                 &      &                 &                 \\  
            & $3_{12}-3_{03}$  &  46.6                  & 1.5            &                 &                 &             &         &                 &      &                 &                 \\  
\hline                                                                                                                                                                                                %
HD~163296   & $1_{11}-0_{00}$  & 158.0                  & 9.0            & 2.3             & 9333            & 30.2        & A1Ve    & 0.07            & 122  & 450             & (6,19,30)       \\  
            & $1_{10}-1_{01}$  & 236.0                  & 9.0            &                 &                 &             &         &                 &      &                 &                 \\  
\hline                                                                                                                                                                                                %
LkCa~15     & $1_{10}-1_{01}$  &  20.0                  & 3.0            & 1.05            & 4375            & 0.74        & K5      & 0.03            & 145  & 900             & (3,9,14,26,31)  \\  
            & $3_{12}-2_{21}$  & 675.0                  & 3.0            &                 &                 &             &         &                 &      &                 &                 \\  
            & $3_{12}-3_{03}$  &  89.0                  & 3.0            &                 &                 &             &         &                 &      &                 &                 \\  
\hline                                                                                                                                                                                                %
MWC~480     & $1_{10}-1_{01}$  &  34.5                  & 5.0            & 2.2             & 8710            & 32.4        & A3ep+sh & 0.04            & 131  & 170             & (7,21)          \\  
            & $3_{12}-2_{21}$  &1125.0                  & 5.0            &                 &                 &             &         &                 &      &                 &                 \\  
            & $3_{12}-3_{03}$  & 140.0                  & 5.0            &                 &                 &             &         &                 &      &                 &                 \\  
\hline                                                                                                                                                                                                %
AS~209      & $1_{10}-1_{01}$  &  75.3                  & 3.0            & 0.9             & 4250            & 1.5         & K5      & 0.028           & 125  & 120             & (1,2,15,32)     \\  
BP~Tau      & $1_{10}-1_{01}$  &  52.1                  & 2.0            & 0.77            & 4055            & 0.83        & K7      & 0.0012          & 56   & 120             & (3,10,13,18,33) \\  
GG~Tau      & $1_{10}-1_{01}$  &  86.0                  & 3.0            & 0.12            & 3055            & 0.065       & M5.5    & 0.01            & 140  & 500             & (3,16,22,34)    \\  
GM~Aur      & $1_{10}-1_{01}$  &  75.0                  & 3.0            & 1.22            & 4750            & 1.01        & K3      & 0.04            & 140  & 300             & (1,3,9,24,35)   \\  
MWC~758     & $1_{10}-1_{01}$  &  74.6                  & 3.0            & 1.8             & 7600            & 11          & A8Ve    & 0.01            & 200  & 250             & (8,12,25)       \\  
T~Cha       & $1_{10}-1_{01}$  &  85.6                  & 4.0            & 1.1             & 5888            & 1.35        & G2:e    & 0.001           & 66   & 230             & (7,20)          \\  
\hline
\end{tabular}
\tablenotetext{a}{Calculated from $M_\text{dust}$ assuming a dust-to-gas mass ratio of 0.01.}
\tablenotetext{b}{These are positive detections, not upper limits.}
\tablerefs{
1. \citet{Herbig1988};
2. \citet{Andrews2009};
3. \citet{White2001};
4. \citet{Andrews2012};
5. \citet{Bergin2013};
6. \citet{vandenAncker1997};
7. \citet{vandenAncker1998};
8. \citet{Beskrovnaya1999};
9. \citet{Furlan2009};
10. \citet{JohnsKrull1999};
11. \citet{Briceno2002};
12. \citet{Chapillon2008};
13. \citet{Simon2000};
14. \citet{Kraus2012};
15. \citet{Andrews2010};
16. \citet{Kenyon1994};
17. \citet{Mulders2013};
18. \citet{Dutrey2007};
19. \citet{Tilling2012};
20. \citet{Huelamo2015};
21. \citet{Hamidouche2006};
22. \citet{Dutrey1997};
23. \citet{Panic2009};
24. \citet{Schneider2003};
25. \citet{Isella2010};
26. \citet{Pietu2006};
27. \citet{Cox2013};
28. \citet{Leinert2004};
29. \citet{Guilloteau1994};
30. \citet{Grady2000};
31. \citet{Isella2012};
32. \citet{Huang2016};
33. \citet{Dutrey2003};
34. \citet{Kawabe1993};
35. \citet{Hughes2009};
}
\end{table*}


\section{A grid of models}
\label{secModel}

We aim to gain insights into the disk chemistry by a quantitative comparison
between the observed upper limits of the water lines with predictions from
chemical models.
The code for this work is the same as that used in \citet{Du2014} and
\citet{Du2015}.  The disk surface density profile is modeled with an analytical
prescription (see, e.g., \citealt{Andrews2009})
\[
\Sigma\propto \left(\frac{r}{r_\text{c}}\right)^{-\gamma} \exp\left[-\left(\frac{r}{r_\text{c}}\right)^{2-\gamma}\right],
\]
where $\gamma$ is fixed to 1.5, and $r_\text{c}$ is fixed to 400~AU (hence
essentially a power-law profile).  This profile is fixed to limit the number of
models in the grid.  The power-law index $\gamma$ in the above formula does not
affect the predicted line intensities significantly (when taking values in the
usual range $\sim1-2$).  The vertical structure is calculated based on
hydrostatic equilibrium, iteratively determined by calculating dust temperature
with Monte Carlo radiative transfer (\citealt{Dullemond2004a}).  The gas
temperature is calculated based on heating-cooling balance.  The chemical
calculation is based on the UMIST 2006 network, extended with adsorption,
desorption, and dust surface chemistry.

We run a grid of models for a range of values for four key parameters of a
protoplanetary disk (\reftab{tabParams}).  We assume that for each disk, its
dust mass is well-determined, but its gas mass is not.  For a fixed dust mass
of the disk, a lower dust-to-gas mass ratio means a higher gas mass.  To
simulate the effect of loss of oxygen due to ice settlement and migration, we
introduce an oxygen depletion factor, which is the oxygen abundance in the
model relative to the ISM value.  The gas mass and the degree of oxygen
depletion are partially degenerate in determining the water emission intensity,
but not completely due to their different effect on the temperature calculation.
The disk inner radius is fixed to 4~AU, and the outer radius is fixed to
400~AU.  Test runs show that the exact value of the inner radius does not
significantly affect the water emission under consideration of this work.  The
effect of the disk outer radius will be discussed in later sections.

\begin{table}[htbp]
\centering
\caption{Parameters for the grid of models.}
\label{tabParams}
\begin{tabular}{p{0.4\linewidth}c}
\hline\hline
Stellar type        &   B, A, F, G, K, M    \\
Disk dust mass      &   $(0.25, 0.5, 1.0, 2.0)\times10^{-4}$~$M_\odot$  \\
Dust-to-gas mass ratio  &   0.01, 0.1, 1.0  \\
Oxygen depletion    &   $10^{-6}$, $10^{-4}$, 0.01, 0.1, 1.0  \\
\hline
\end{tabular}
\end{table}

\begin{figure}[htbp]
\includegraphics[width=\linewidth]{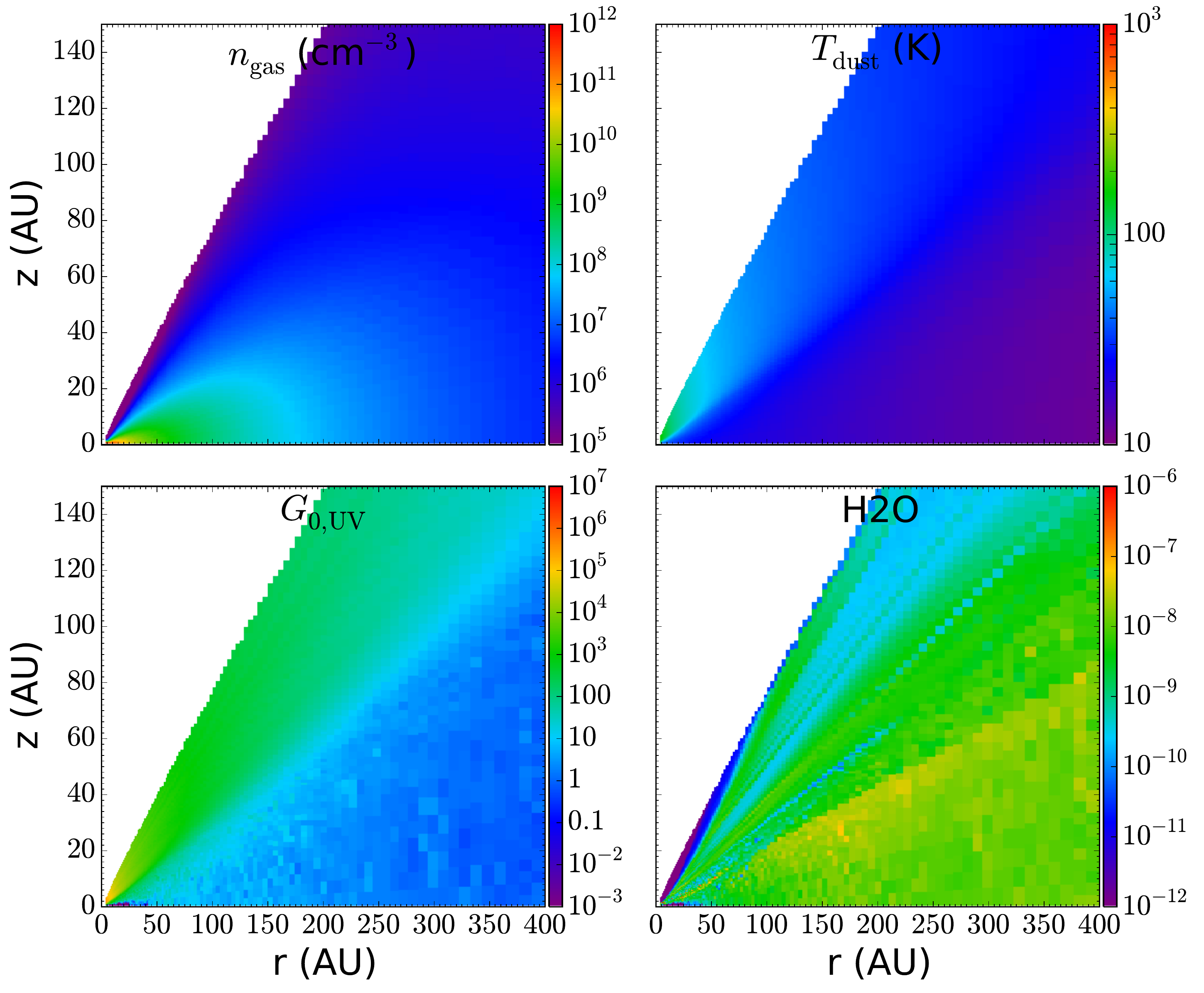}
\caption{Gas density, dust temperature, UV intensity, and water vapor abundance in an example model.  The underlying disk has a dust-to-gas mass ratio of $10^{-2}$, dust mass of \(10^{-4}\) \(M_\odot\), a stellar spectral type of G, and no oxygen and carbon depletion.}
\label{figMod004}
\end{figure}

\section{Results}
\label{secResults}

\subsection{Observed upper limits versus the models}

The correspondence between a specific model in the grid and each of the observed sources is based on matching between the stellar types and the measured disk dust masses.
The observed intensities or upper limits of the water lines for each source are scaled to a distance of 100~pc for comparison.  We further scaled the values for the \Ti{} and \Tii{} lines with a disk outer radius of 400~AU assuming the line emission is uniformly distributed over the whole disk, namely, for each source, the line intensity (upper limit) to be compared with the models is calculated from the observed value by
\begin{equation*}
  I_\text{model} = I_\text{obs} \left(\frac{400\ \text{AU}}{R_\text{out}}\right)^2 \left(\frac{d}{100\ \text{pc}}\right)^2.
\end{equation*} 
We did not scale the intensities of \Tiii{} and \Tiv{} with disk size (but they are still scaled with distance), because they mostly originate from the inner disk (see below).
For a subset of the parameter space, Figs. \ref{fig110101} and \ref{fig312221} show the placement of the observed upper limits (black arrows) for the \Ti{} and \Tiii{} transitions together with some detections relative to the models.
For more complete data, see Figs. \ref{fig110101a}, \ref{fig111000a}, \ref{fig312221a}, and \ref{fig312303a} (in appendix~\ref{appendixFigs}).  In these figures, each panel corresponds to a specific dust-to-gas mass ratio (d2g) and disk dust mass ($m_\text{dust}$).  Each panel shows the modeled line intensity as a function of stellar type with different oxygen depletion factors.

\begin{figure*}[htbp]
\includegraphics[width=\linewidth]{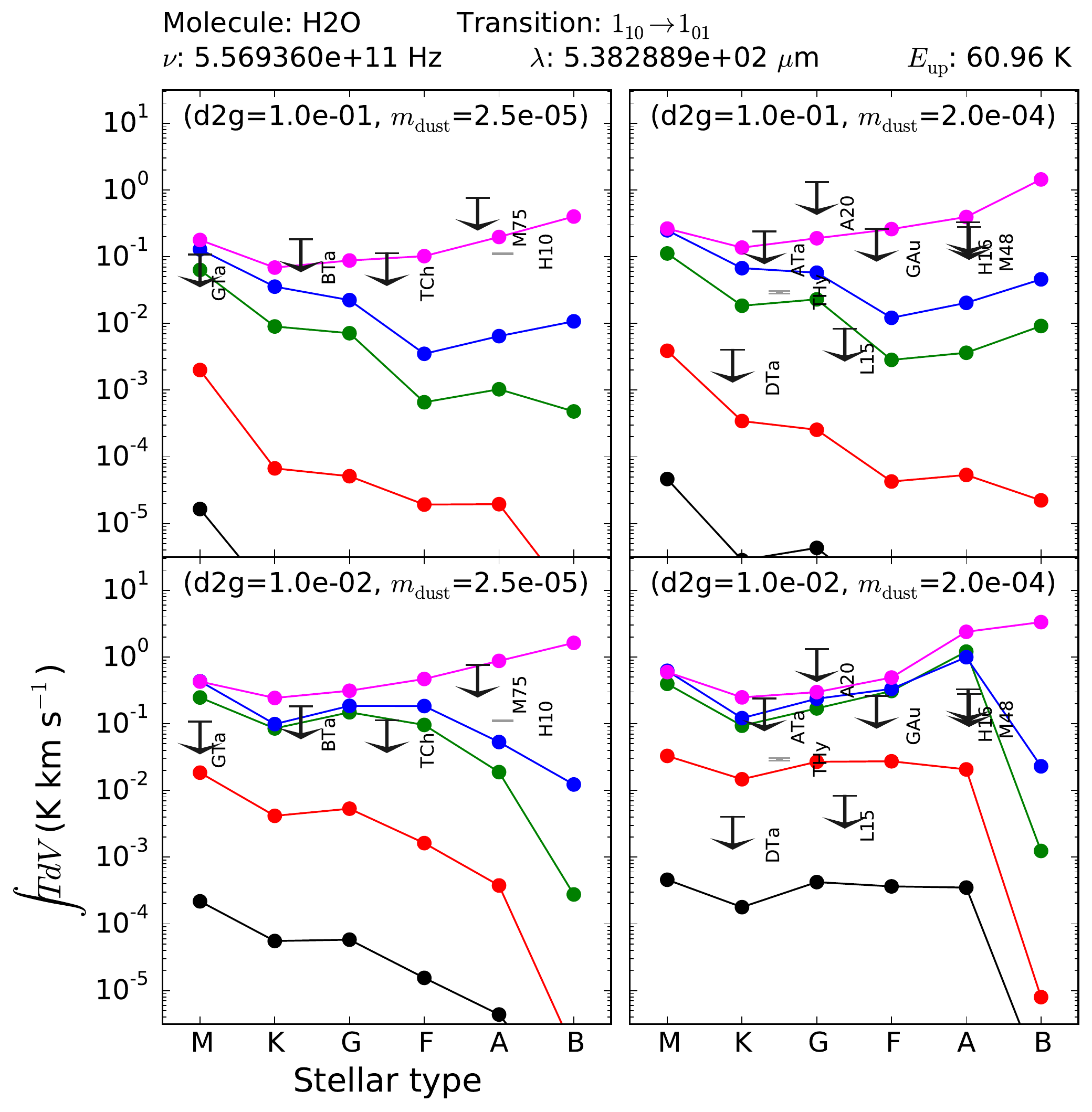}
\caption{The modeled water \Ti{} line intensities (round dots),
together with the observed upper limits (black arrows; scaled to a distance of
100~pc and a disk outer radius of 400~AU) and detections (gray error bars).  Each panel corresponds to a combination of dust-to-gas mass ratio (varying in the vertical direction) and dust mass (varying in the horizontal direction).  The abbreviated source names are shown for each data point (upper limit or error bar); e.g., HD~100546 is shown as H10, and GG~Tau as GTa (cf. \reftab{tabObsData}).  The upper limits are three times
the RMS noise plus 20\% of systematic uncertainty (the head of each arrow is
located at one third of each upper limit).  Different color of the
models means different degree of oxygen depletion. Magenta: no oxygen
depletion; blue: 0.1; green: 0.01; red: $10^{-4}$; black: $10^{-6}$.}
\label{fig110101}
\end{figure*}

\begin{figure*}[htbp]
\includegraphics[width=\linewidth]{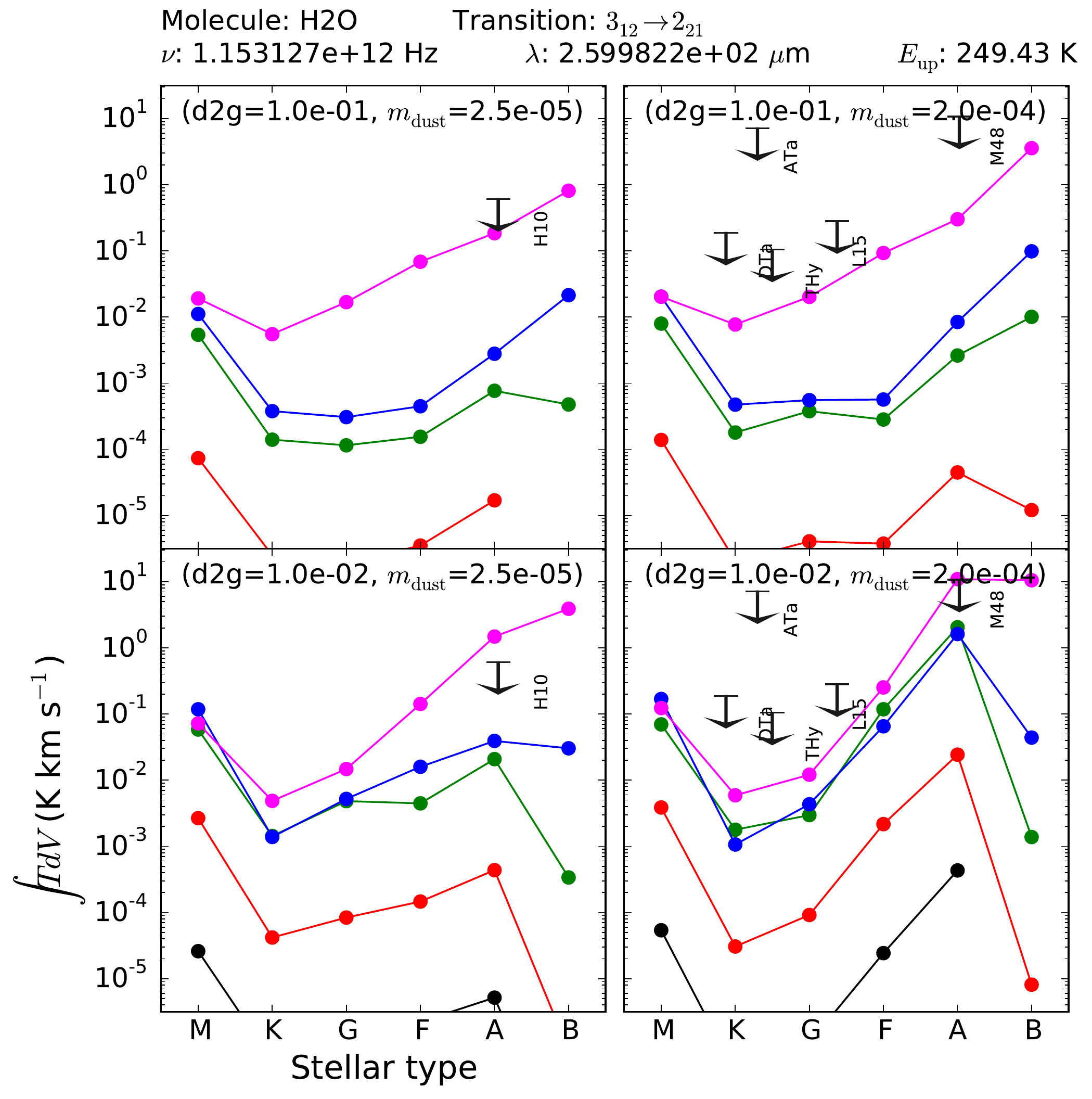}
\caption{The same as \reffig{fig110101} except that it is for the \Tiii{} line.}
\label{fig312221}
\end{figure*}

The \Ti{} and \Tii{} lines have upper energy levels of 50--60~K (\reffig{fig110101} and \reffig{fig111000a}).  With a normal dust-to-gas mass ratio (0.01; bottom panels in the figures), they require a high degree of oxygen depletion (by a factor ${\sim}10^{-2}$ to ${<}10^{-4}$) to be consistent with the observed upper limits.  Even in the extreme case of dust-to-gas mass ratio being one (i.e. very low gas mass), the models still tend to over-predict the two lines for most sources by at least a factor of a few, and models with oxygen depletion agree better with the data.

\begin{figure}[htbp]
\includegraphics[width=\linewidth]{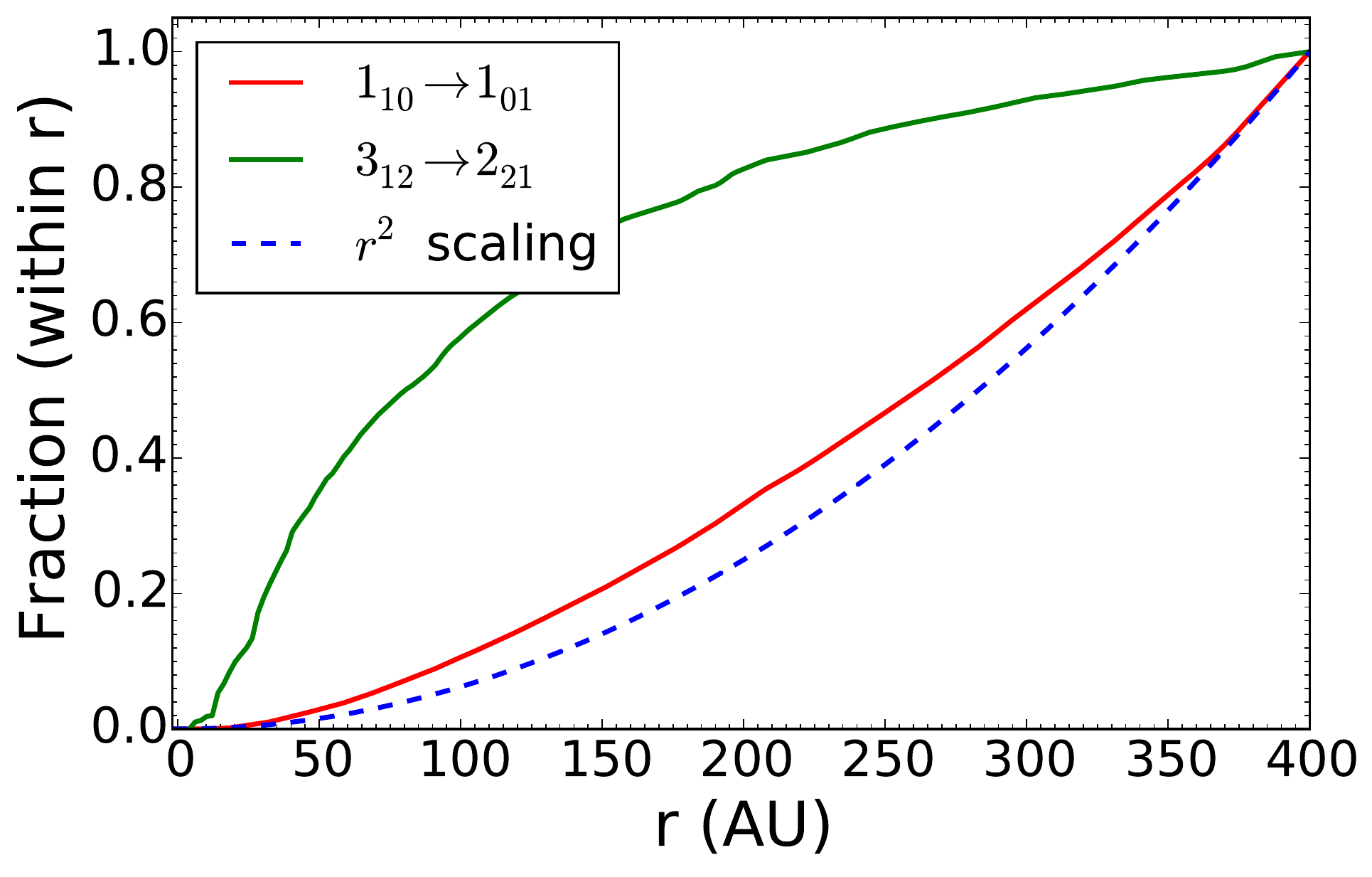}
\caption{The cumulative distribution of emission as a function of radius (i.e. the fractional amount of emission within $r$) for the \Ti{} and \Tiii{} lines.  The underlying disk model is the same as in \reffig{figMod004}.  The dashed line shows a $r^2$-scaling relation, which is used for scaling the \Ti{} and \Tii{} transitions.}
\label{figContri}
\end{figure}

The observational upper limits of the \Tiii{} and \Tiv{} lines are not as constraining as the other two lines, due to their shallower integration times.  They are consistent with (but do not infer) lower degree of depletion for oxygen or no depletion at all.  It is possible that future observations may provide a more stringent limits on their fluxes, which will require the depletion of oxygen to a degree similar to the other two lines.  On the other hand, the upper state energies of \Tiii{} and \Tiv{} are $\sim$250~K, so they mostly originate from the inner warm part of the disk.  This can be seen in \reffig{figContri}, which shows the accumulative distribution of the \Ti{} and \Tiii{} lines as a function of radius.  A low level of oxygen depletion in the inner disk is consistent with (or even needed for) a scenario found in \citet{Du2015} (see also \citealt{Bergin2016}), in which the elemental abundance of oxygen in the inner disk is not as depleted as in the outer disk, possibly due to inward migration and subsequent evaporation of icy dust particles.

\subsection{Effect of the disk size}
\label{subsecDiskSize}

In our grid of models the disk outer radius is fixed to 400~AU, and for comparison we scaled the observational intensities or upper limits for \Ti{} and \Tii{} to this disk size, using disk sizes found in the literature.  Such a scaling is reasonable, since these two lines are roughly uniformly distributed in the outer disk (cf. \reffig{figContri}).  One issue with this procedure is that for many disks their sizes are not well-known, or not well-defined.  For example, For LkCa~15, the disk outer radius inferred from dust is 150~AU, while \(^{12}\)CO gives a value of 900~AU \citep{Isella2012}.  For the calculation here we always adopt the size of the gas disk as seen in CO when available, otherwise we use the size inferred from scattered light.  \citet{Salinas2016} and Hogerheijde et al. (in prep) also address the radial location of the detected low-excitation water emission lines in TW Hya and HD~100546.  It is important to get reliable estimates of $r_\text{in}$ and $r_\text{out}$.

For the \Tiii{} and \Tiv{} lines the disk size does not play a significant role, since they mostly originate from the inner warm disk, so we did not rescale them for the disk size.

One may say that water vapor does not necessarily coexist with gas tracers such as CO, or, in other words, the sizes of the water vapor disks may likely be much smaller than that of the CO disks.  Actually, this is exactly our point: in the outer disk the abundance of gas-phase oxygen (hence water vapor) must be much lower than model predictions from canonical elemental abundances to match the data, which is equivalent to saying that the oxygen (and water vapor) distribution is radially much more compact than other species (such as H$_2$).  One limitation of the current grid of models is that we assume (to limit the number of variables) the elemental abundances are uniform over the whole disk; in more detailed models for TW~Hya and DM~Tau as described in \citet{Du2015} and \citet{Bergin2016}, we found that nonuniform elemental distributions (in which oxygen and carbon become more abundant towards the inner disk) are needed to match well with richer datasets (more lines from water and from other species), which is essentially saying that the disk size as seen in volatile oxygen (and carbon) is small compared to the overall gas disk.

\subsection{Detection of the \Ti{} line in stacked spectra of AA~Tau, DM~Tau, LkCa~15, and MWC~480}
\label{secStackDetec}

\begin{figure}[htbp]
\includegraphics[width=\linewidth]{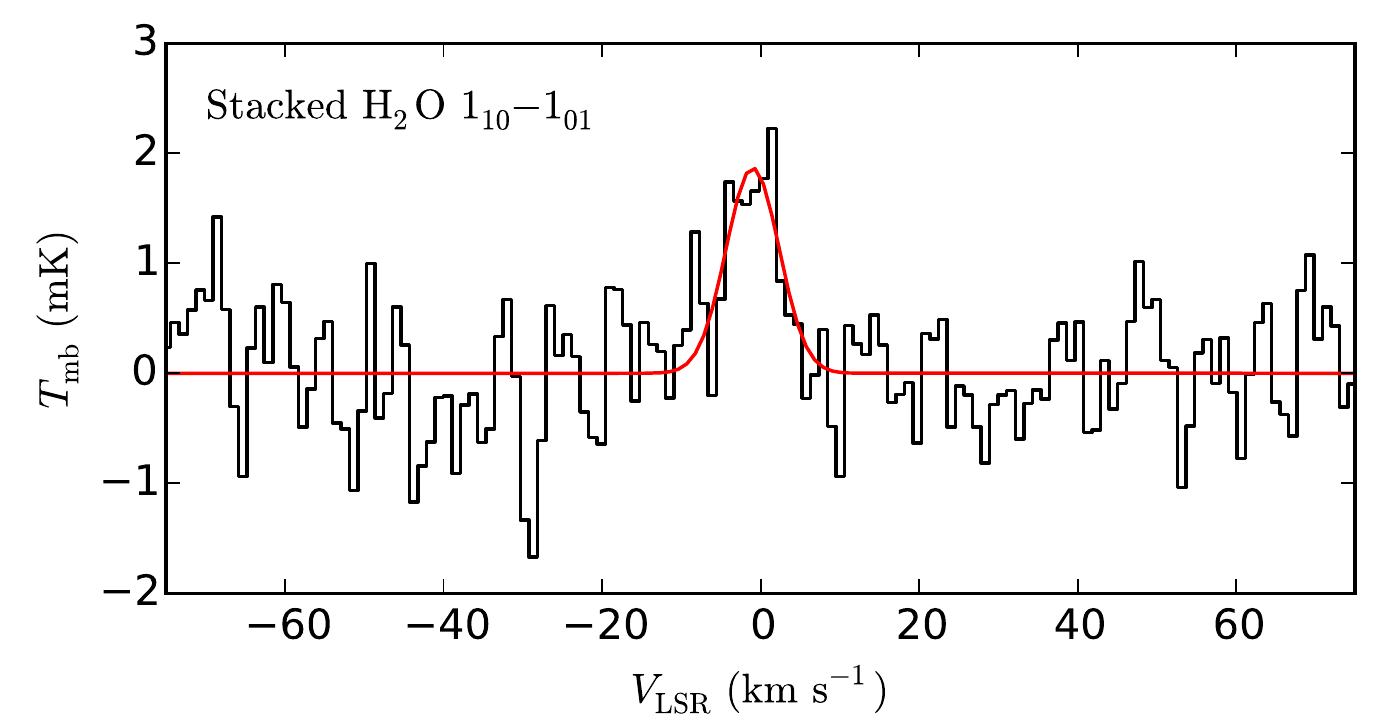}
\caption{Detection of the water $1_{10}-1_{01}$ line (557 GHz) in the stacked
spectra of AA~Tau, DM~Tau, MWC~480, and LkCa~15.  The integrated intensity
(weighted average) is 15$\pm$3 mK~km~s$^{-1}$.}
\label{figDetection}
\end{figure}

The \Ti{} line is not detected in the individual spectrum of AA~Tau, DM~Tau, LkCa~15, and MWC~480 (the deepest integrations in our survey), but when their spectra are stacked together after correcting for their different velocities, a feature with an integrated intensity of ${\sim}5\sigma$ emerges (\reffig{figDetection}).  Weighted stacking including all the sources gives similar result.  Assuming this line has roughly the same intensity in these four disks, then its intensity in each of them would be 15${\pm}3$~mK~km~s$^{-1}$.  Comparing this value with \reffig{fig110101} shows that, in the case of dust-to-gas mass ratio of 0.01, oxygen has to be depleted relative to the ISM value by a factor of $\sim10^{-4}$ (at least in the upper emissive layers) to match the detected intensity.

\section{Discussions and implications}
\label{secDiscussion}

We have surveyed four water lines (two ground state transitions and two at higher levels, with the former two conducted with very deep integration and the latter two with shallower integration) in 13 protoplanetary disks.  With the exception of TW~Hya, HD~100546, and the stacked data of AA~Tau, DM~Tau, LkCa~15, and MWC~480, the lines are not detected at the present noise level, while a higher detection rate for the ground state lines is expected from canonical models.

Previously it has been observationally found that the low-energy emission of water, O~\textsc{i}, and CO in TW~Hya tend to be weaker than values expected from thermo-chemical models.  The current work shows that this is true in a larger sample, with typical degree of depletion for oxygen of the order of $10^{-2}$ to $10^{-4}$.  In our view, the most likely underlying mechanism for this is the coupled evolution of dust and gas chemistry \citep{Bergin2016}, in which the dust particles act as a sink of the volatiles and a vehicle to transport them to the midplane and the inner disk.  An analytical study based on similar ideas can be found in \citet{Kama2016a}, and a more detailed Monte Carlo simulation is given by \citet{Krijt2016}.  The low water line emission intensity in the sample indicates that this is likely a ubiquitous phenomenon.  Recently \citet{Antonellini2016} proposed the high noise level caused by the continuum flux as the most likely explanation for the low detection rate of mid-IR water lines in disks around Herbig stars; in our case, the continuum flux is much weaker, and the descrepancy between the upper limits from the data and the canonical models provide useful information about the chemical and physical processes in the disks, also for Herbig stars.

The depletion of volatiles may not be uniform throughout the disk or be of the same degree for different sources (see, e.g., \citealt{Ciesla2006}).  The water emission lines in this work originate from cold or lukewarm water vapor (with upper state energies 53 and 249~K), as opposed to the hot water vapor emission lines detected by \citet{Salyk2008} and \citet{Pontoppidan2010b}.  Water vapor would exist throughout the disk, and in the outer disk below its freezing temperature it is released from the dust grains by photo-desorption \citep{Walsh2010}.  The current work, together with previous related studies, shows that cold water vapor is depleted (i.e. lower than expected from canonical models).

For the inner disk, a number of studies have found very high column
densities of hot water vapor, from $10^{17}$--$10^{21}$~cm$^{-2}$, with values
depending on the assumed emitting area (e.g., \citealt{Salyk2008} and
\citealt{Doppmann2011}). Abundance ratios of water relative to CO of
1-10 generally indicate high water abundances of order $10^{-4}$--$10^{-5}$,
suggesting that the disk surfaces are not ``dry'' (e.g., \citealt{Salyk2011}, \citealt{Mandell2012}), although some classes of disks
appear to have less water in their innermost regions (e.g., \citealt{Najita2013}, \citealt{Zhang2013}, \citealt{Banzatti2017}).  \citet{Carr2004} found a factor of 10 depletion for the hot ($\sim$1500~K) water vapor within 0.3~AU of the young stellar object SVS~13.  This value is consistent with \citet{Krijt2016}, but is still much less than the typical values found in the present survey.
\citet{Blevins2016} demonstrate a quantitative drop of at least 5 orders of
magnitude in water abundance from inner to outer disk for their
sample of disks with deep VLT, Spitzer and Herschel data, similar to
what is found here.

The prominent difference between the cold and hot water vapor indicates that the outer disk and inner disk are subject to different processes.  While dust grains are growing and settling down to the midplane in the outer disk, in the inner disk they might be stirred up by certain mechanisms and return volatiles back to the gas phase (interior to the snowline even large bodies may evaporate).  Due to the differential nature of dust sedimentation, icy particles with small size can continue to exist in the outer disk for a long time, giving rise to water ice features \citep{Chiang2001}.

A low gas-to-dust mass ratio (maybe close to or even lower than one) could also explain the non-detections.  
Disks lose gas and dust through accretion and photoevaporation \citep{Hartmann1998,Adams2004,Alexander2006,Owen2012,Bae2013,Gorti2015}, and possibly through other mechanisms \citep{Hollenbach2000,Alexander2014}.
The mass loss rate due to photoevaporation typically lies in the range \(10^{-10} - 10^{-7}\) \(M_\odot\) \citep{Owen2011}.
In photoevaporating flows, small dust particles are entrained by gas, while large dust particles are left behind \citep{Hutchison2016}.
\citet{Gorti2009} found that FUV photoevaporation can deplete most of the mass of a 0.03 \(M_\odot\) disk with timescales \({\sim}1\) Myr, and is most effective in the outer disk.
\citet{Gorti2015} found that \({\sim}3\times10^{-4}\) \(M_\odot\) of mass in solids remain after gas disk dispersal.
At face value this gives a low gas-to-dust mass ratio.

However, the low value definitely does not apply to sources with gas mass determined directly \citep{Bergin2013,McClure2016}.
We also note that the outer gas disks detected in CO or scattered light for some sources in our sample are known to be not completely gone, and extend to a few hundred AU.  The observational detection of \ce{N2H+} and \ce{HCO+} \citep{Pietu2007,Qi2013b,Thi2004} in the outer disk also requires the very existence of hydrogen gas (see \citealt{Aikawa2015} for detailed calculations), and the CO emission in the outer disk also means some amount of gas still exist there.
Furthermore, some disks (e.g. DM~Tau, AS~209, and MWC~480) in our sample are still actively accreting.  For an accretion rate of \({\sim}10^{-8}\) \(M_\odot\) yr\(^{-1}\) \citep{Hartmann1998,JohnsKrull2000,Grady2010}, if the gas mass is indeed of the order of \(10^{-4}\) \(M_\odot\), then the disk would be gone within \({\sim}10^{4}\) yr, or, in other words, the probability of seeing the disk in its current state would be \({\lesssim}0.01\).
In reality, it is likely that chemical depletion coupled with dust evolution as we describe here and elsewhere \citep{Du2015,Bergin2016} occurs with a timescale shorter than that of photoevaporation (see, e.g., \citealt{Dullemond2005a}), perhaps even in the deeply embedded phases \citep{Anderl2016}.

As discussed extensively in \citet{Bergin2016}, the depletion of oxygen can have a large effect on the abundances of other species, especially the hydrocarbons \citep{Du2015}.  When oxygen is at its canonical ISM abundance (\(\sim3\times10^{-4}\)), most of the carbon atoms end up in CO.  But when oxygen is depleted relative to carbon, the abundances of hydrocarbons will be significantly enhanced.  Hence it would be interesting to survey hydrocarbon emission in these sources.  The creation of hydrocarbons is also tied with dust evolution, in that in a segregated disk, with larger-sized grains mostly being in the inner region and smaller-sized grains more diffusively distributed, the UV photons will be able to penetrate deeper in the outer disk, to create large amounts of hydrocarbons.  This process is also an evolutionary effect, in the sense that disks with different ages will show different degrees of volatile depletion.  Efforts are ongoing to search for hydrocarbon emission in these (but not limited to) sources to test this prediction.

\section{Summary}
\label{secSummary}

In this paper we present a survey of four water lines (two of them with very deep integration) in 13 protoplanetary disks, and compare the observed results with a grid of models.  The main findings are:
\begin{itemize}
    \item The detection rate is low: only the very-deeply integrated line(s) are detected in TW~Hya, HD~100546 (not presented in this paper), and in the stacked spectrum of AA~Tau, DM~Tau, LkCa~15, and MWC~480.
    \item To be consistent with the observational detections and upper limits, it is very likely that oxygen is depleted in the emissive layers of the disk (especially in the outer disk) by a factor of $\sim$100--$10^4$.
    \item Oxygen distribution in the inner disk is less constrained; a lower depletion of oxygen in the inner disk than in the outer disk is consistent with current upper limits.
\end{itemize}

\acknowledgments
HIFI has been designed and built by a consortium of institutes and university departments from across Europe, Canada and the United States (NASA) under the leadership of SRON, Netherlands Institute for Space Research, Groningen, The Netherlands, and with major contributions from Germany, France and the US.
Support for this work was provided by NASA through an award issued by JPL/Caltech.
EAB acknowledges support from NASA XRP grant NNX16AB48G.
MH and EvD acknowledge support from the Netherlands Research School for Astronomy (NOVA) and European Union A-ERC grant 291141 CHEMPLAN.
Davide~Fedele acknowledges support from the Italian Ministry of Education, Universities and Research project SIR (RBSI14ZRHR).

\bibliographystyle{aasjournal}

\bibliography{references}

\appendix

\section{Modeled intensities of the lines versus observations.}
\label{appendixFigs}

\begin{figure*}[htbp]
\includegraphics[width=\linewidth]{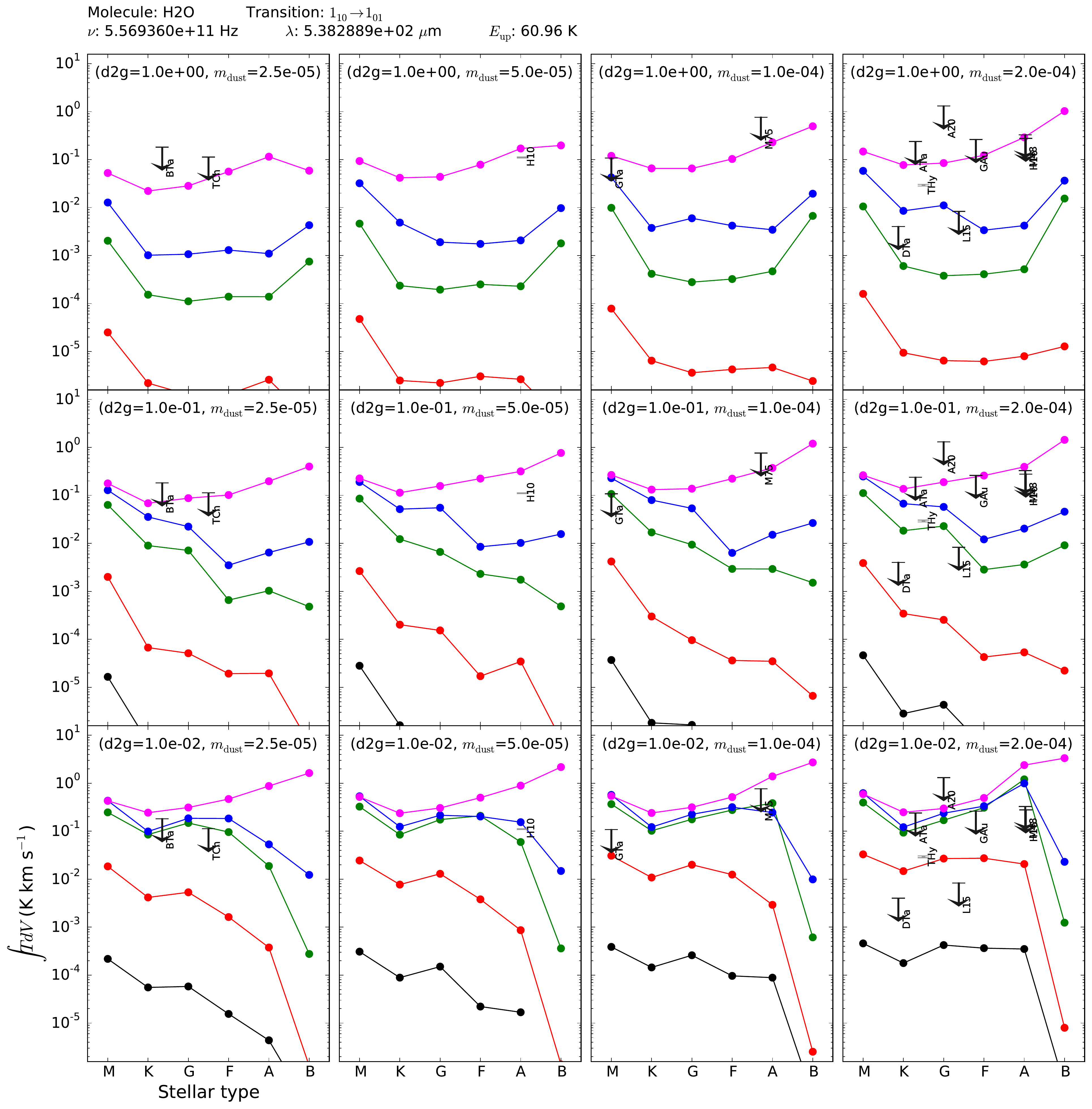}
\label{fig110101a}
\caption{The same as \reffig{fig110101} except that this one includes the complete parameter space of the models.}
\end{figure*}

\begin{figure*}[htbp]
\includegraphics[width=\linewidth]{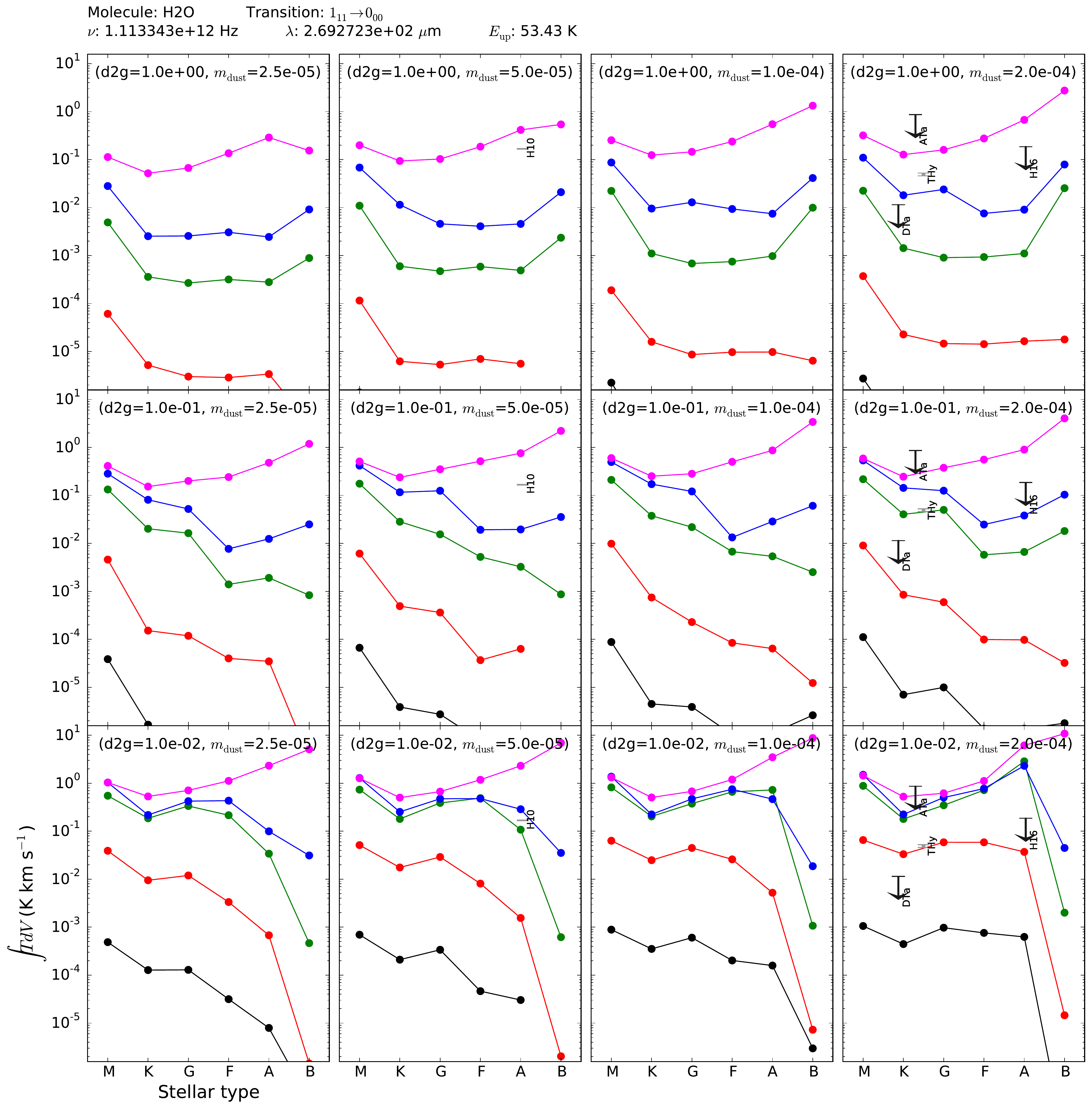}
\caption{The same as \reffig{fig110101a} except that it is for the \Tii{} line.}
\label{fig111000a}
\end{figure*}

\begin{figure*}[htbp]
\includegraphics[width=\linewidth]{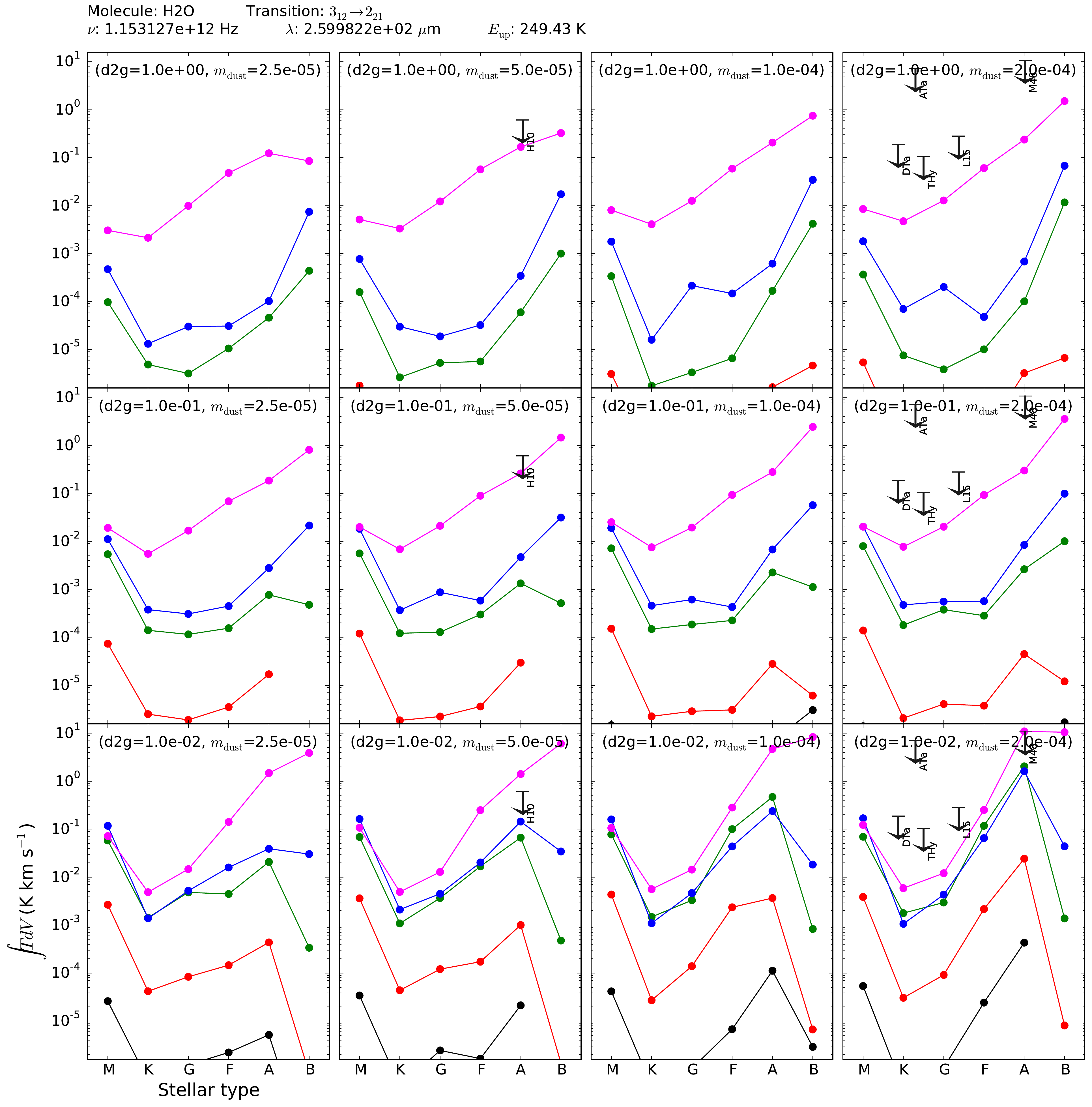}
\caption{The same as \reffig{fig110101a} except that it is for the \Tiii{} line; the difference is that the observed upper limits are \emph{not} rescaled with the disk size.}
\label{fig312221a}
\end{figure*}

\begin{figure*}[htbp]
\includegraphics[width=\linewidth]{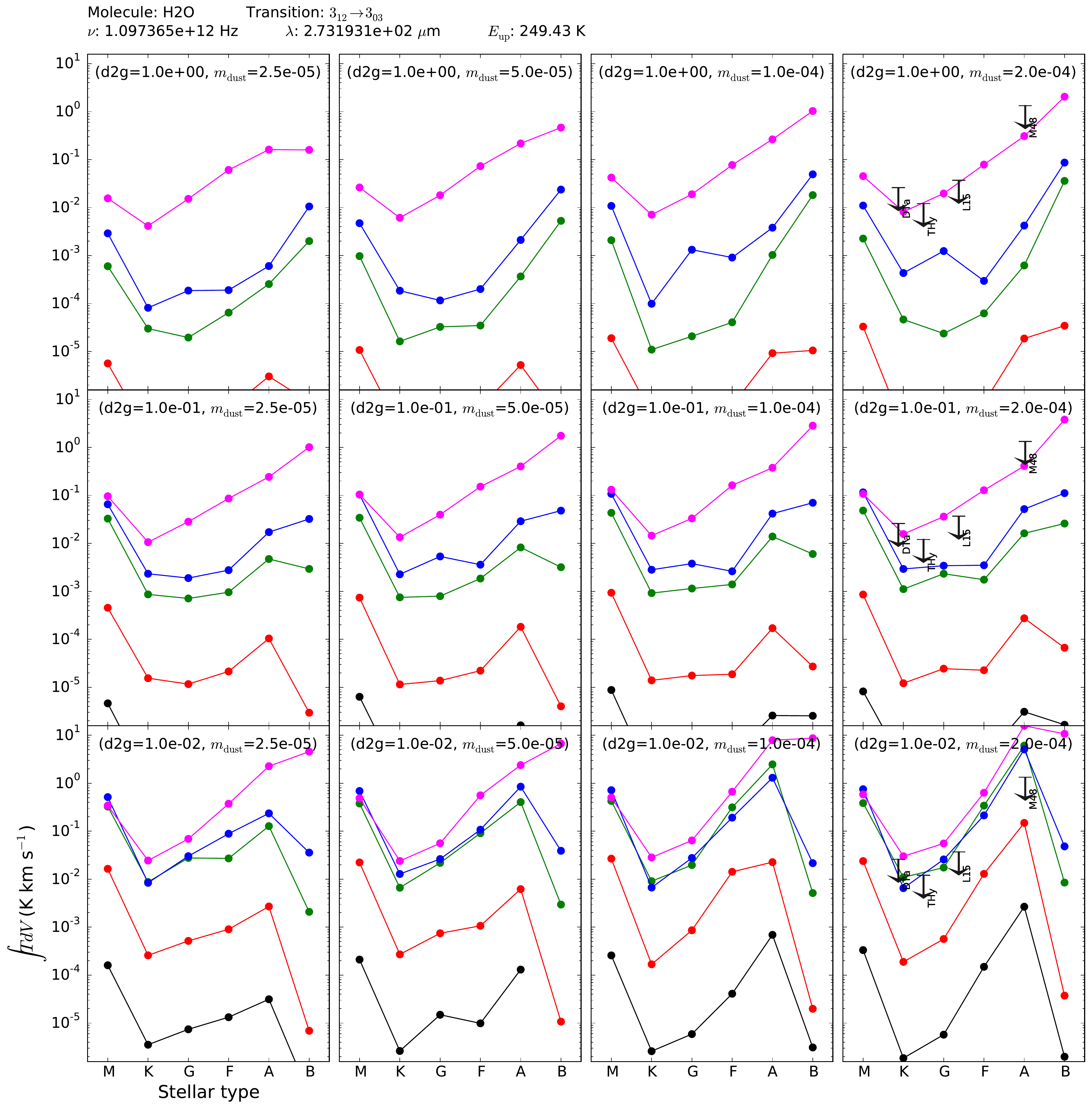}
\caption{The same as \reffig{fig312221a} except that it is for the \Tiv{} line.}
\label{fig312303a}
\end{figure*}

\section{Observed spectra for all sources}
\label{appendixObsFigs}

\begin{figure*}[htbp]
\includegraphics[width=\linewidth]{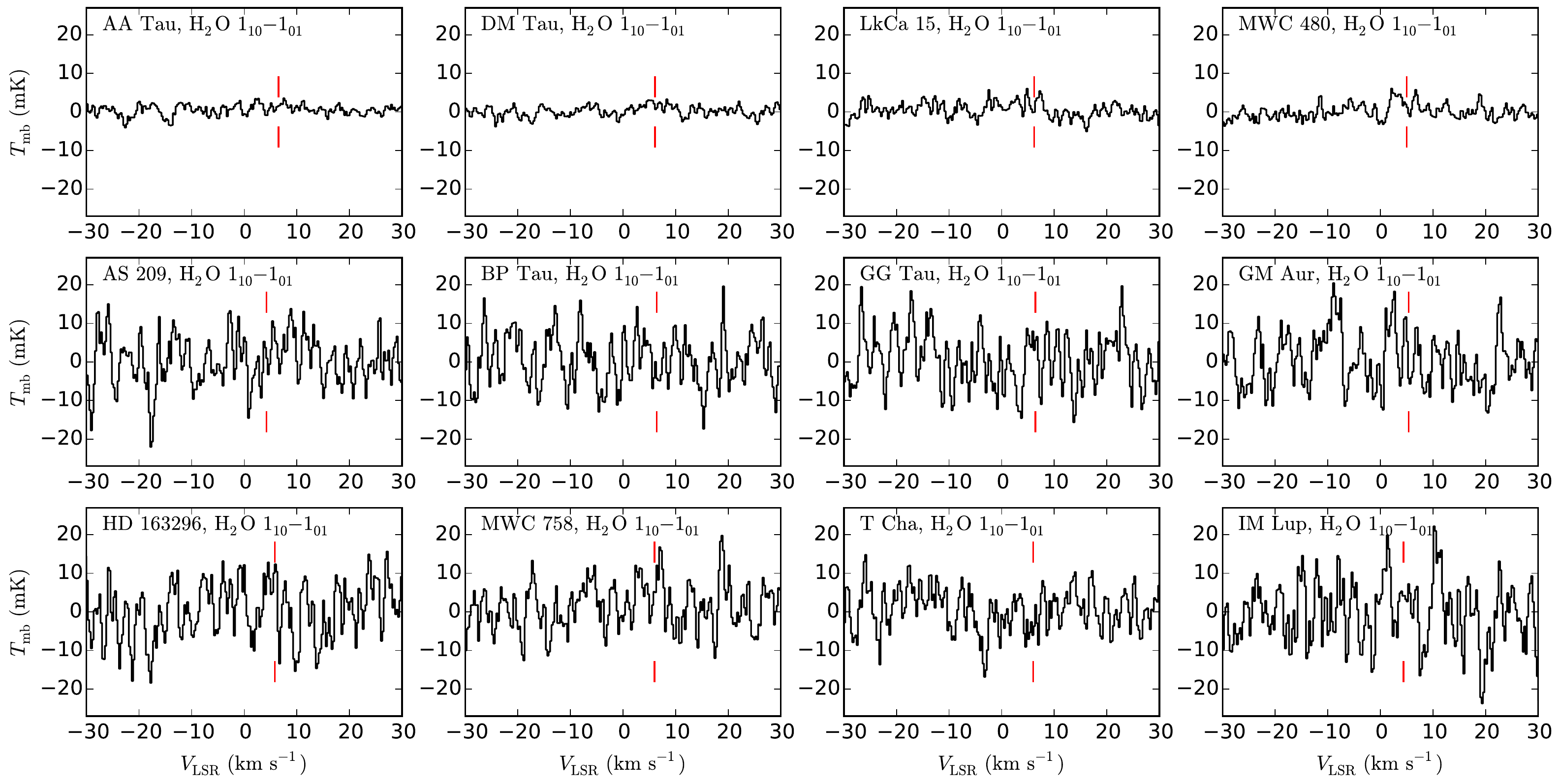}
\caption{The water \Ti{} spectra.  The red line marks the
$V_\text{LSR}$ of each source.}
\label{figTiAll}
\end{figure*}

\begin{figure*}[htbp]
\includegraphics[width=\linewidth]{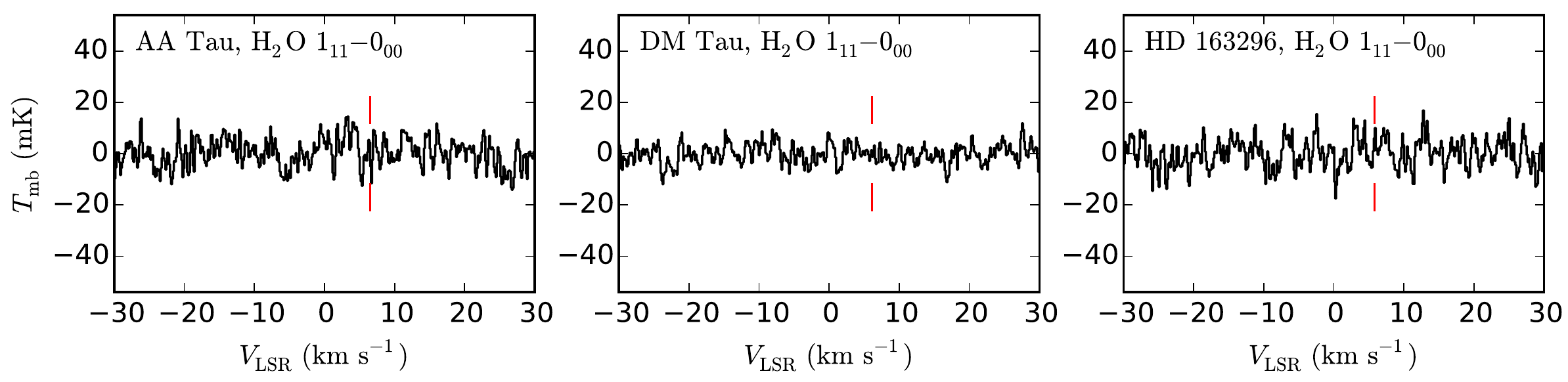}
\caption{The water \Tii{} spectra.}
\label{figTiiAll}
\end{figure*}

\begin{figure*}[htbp]
\includegraphics[width=\linewidth]{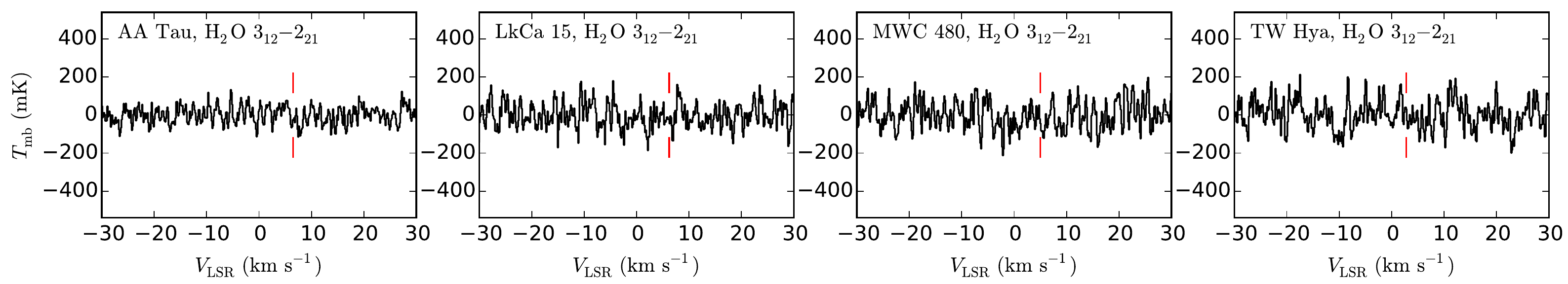}
\caption{The water \Tiii{} spectra.}
\label{figTiiiAll}
\end{figure*}

\begin{figure*}[htbp]
\includegraphics[width=\linewidth]{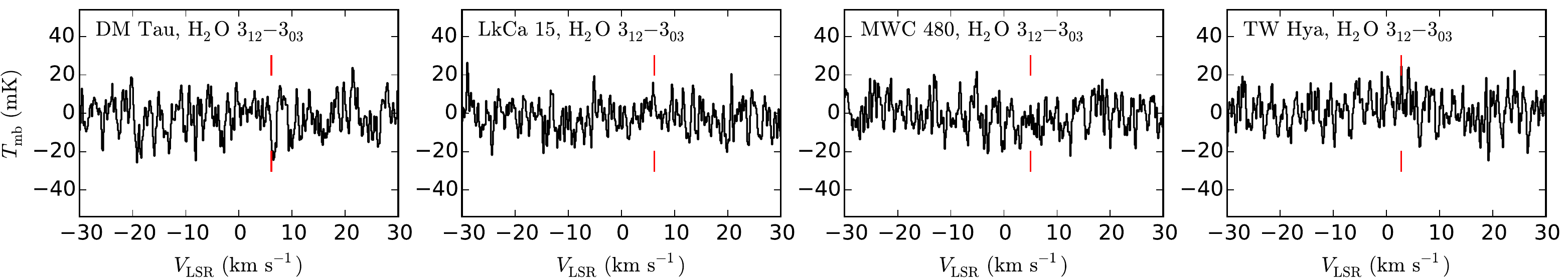}
\caption{The water \Tiv{} spectra.}
\label{figTivAll}
\end{figure*}

\begin{figure*}[htbp]
\includegraphics[width=\linewidth]{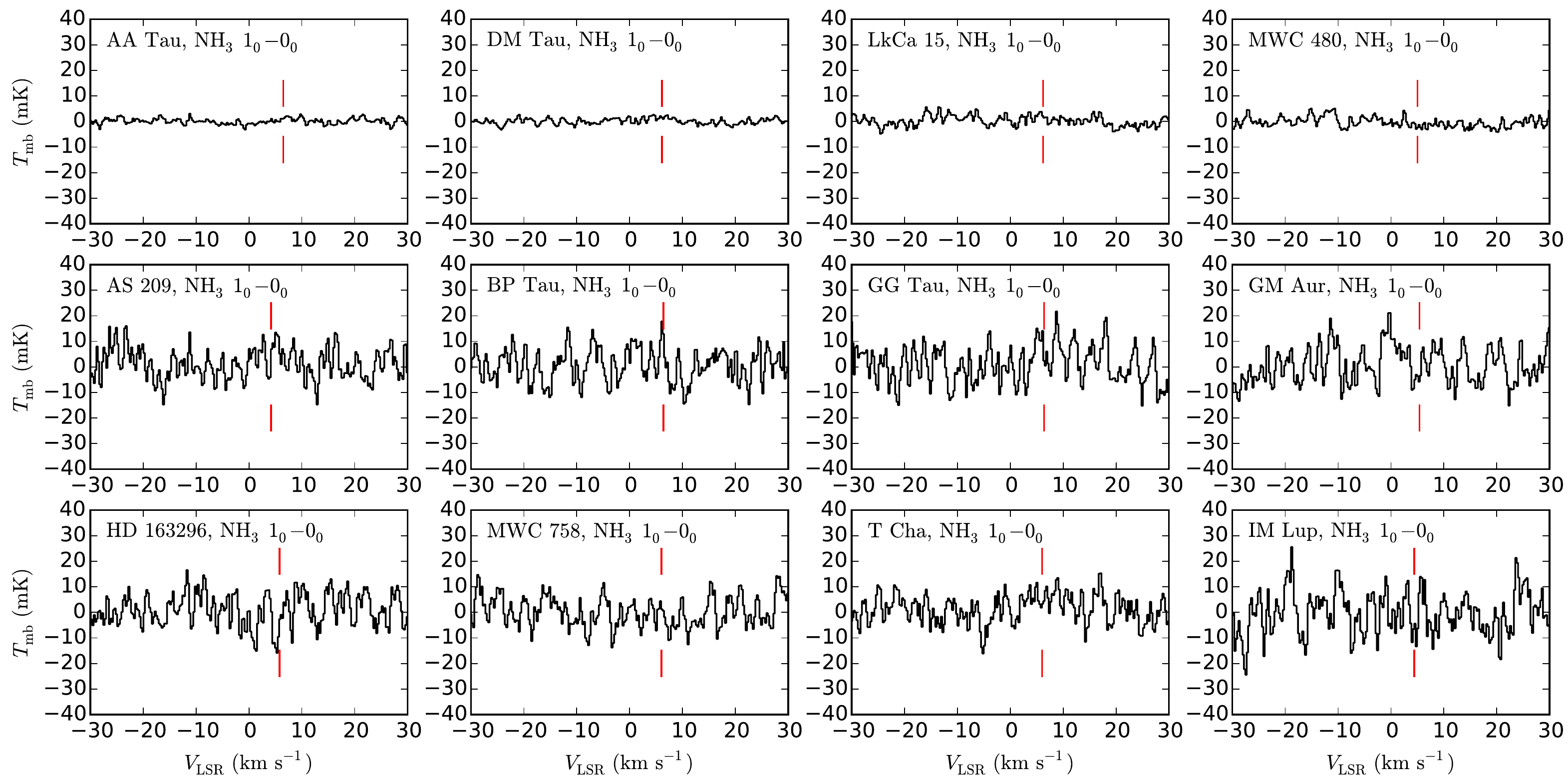}
\caption{The NH$_3$ ($1_0-0_0$) spectra.}
\label{figNHAll}
\end{figure*}

\begin{figure*}[htbp]
\includegraphics[width=\linewidth]{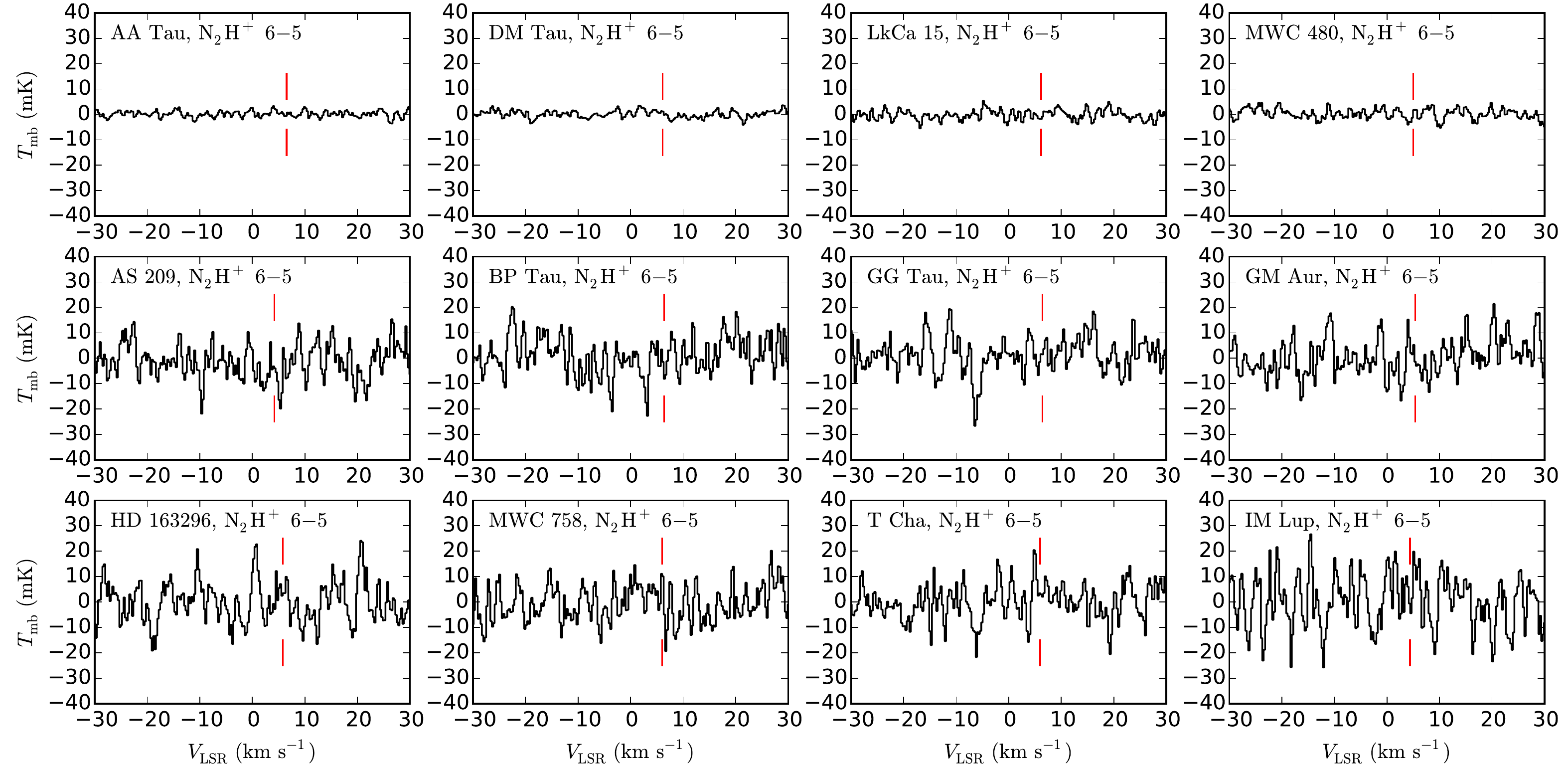}
\caption{The N$_2$H$^+$ (6$-$5) spectra.}
\label{figNtHAll}
\end{figure*}

\begin{figure*}[htbp]
\includegraphics[width=\linewidth]{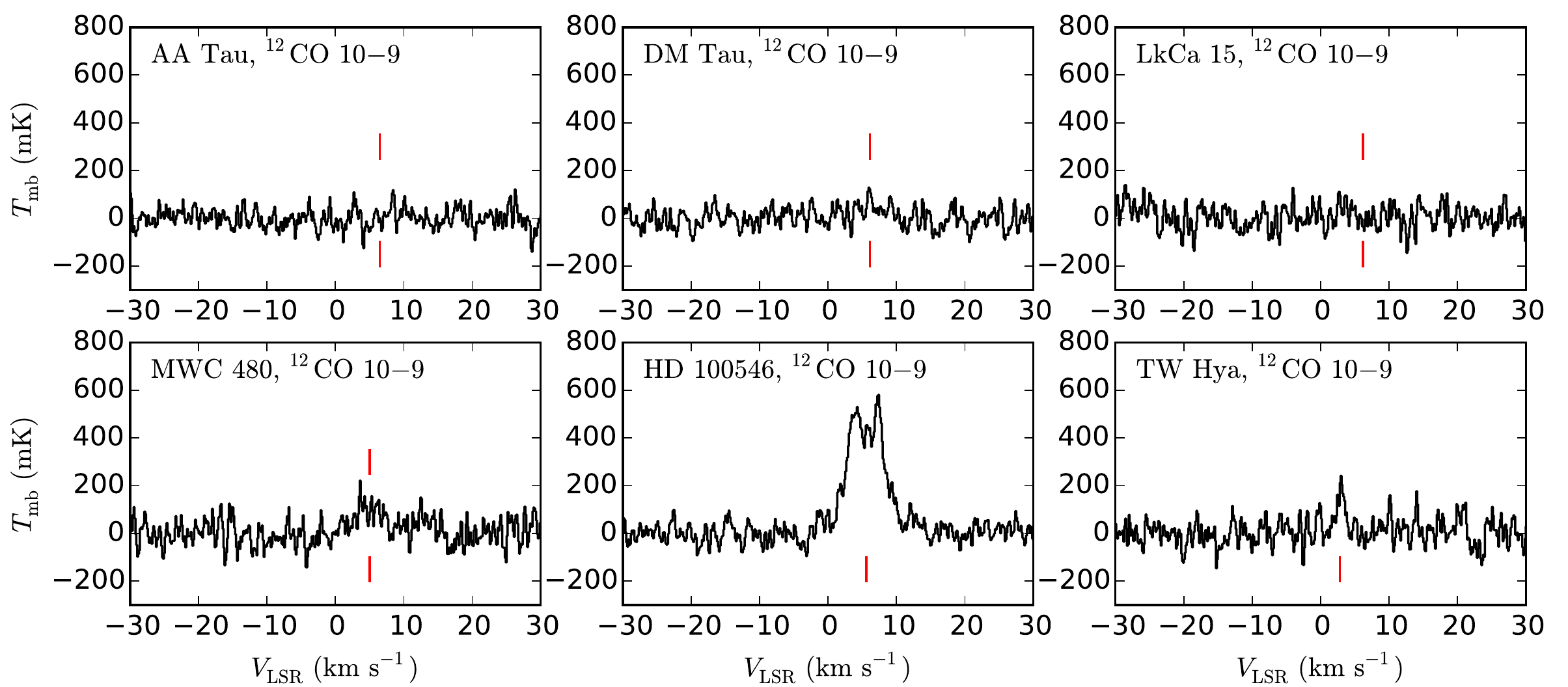}
\caption{The CO (10$-$9) spectra.}
\label{figCOtenAll}
\end{figure*}

\begin{figure*}[htbp]
\includegraphics[width=\linewidth]{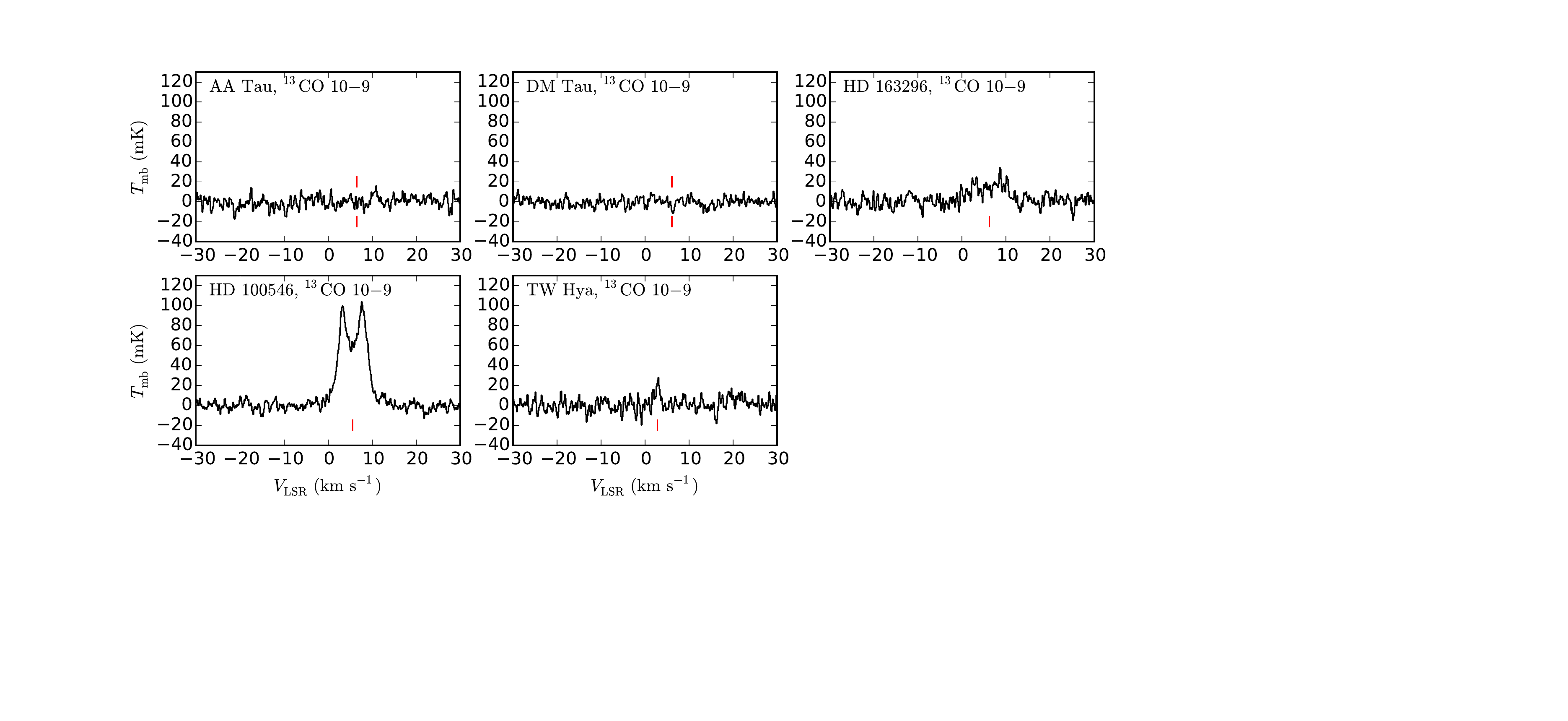}
\caption{The $^{13}$CO (10$-$9) spectra.}
\label{figthCOAll}
\end{figure*}

\end{document}